# Relation between unidirectional spin Hall magnetoresistance and spin current-driven magnon generation

I.V. Borisenko[1, 2,*], V.E. Demidov[1], S. Urazhdin[3], A.B. Rinkevich[4], and S. O. Demokritov[1,4]

[1]*Institute for Applied Physics and Center for Nanotechnology, University of Muenster, 48149 Muenster, Germany*

[2]*Kotel'nikov Institute of Radio Engineering and Electronics, Russian Academy of Sciences, 125009 Moscow, Russia*

[3]*Department of Physics, Emory University, Atlanta, Georgia 30322, USA*

[4]*Institute of Metal Physics, Ural Division of RAS, Yekaterinburg 620108, Russia*

*Abstract.*

We perform electronic measurements of unidirectional spin Hall magnetoresistance (USMR) in a Permalloy/Pt bilayer, in conjunction with magneto-optical Brillouin light spectroscopy of spin current-driven magnon population. We show that the current dependence of USMR closely follows the dipolar magnon density, and that both dependencies exhibit the same scaling over a large temperature range of 80-400 K. These findings demonstrate a close relationship between spin current-driven magnon generation and USMR, and indicate that the latter is likely dominated by the dipolar magnons.

* Corresponding author, e-mail: boriseni@uni-muenster.de

The phenomenon of magnetoresistance (MR) – the dependence of electrical resistance in certain materials on the magnetic field - has found numerous applications in electronics and sensing. For instance, the anisotropic MR (AMR) in magnetic materials[1] has been extensively studied and utilized since its discovery in the 19th century. Developments in thin-film growth have led to the discovery of the giant MR (GMR)[2,3] and tunnelling MR (TMR)[4,5] in thin-film magnetic multilayers that consist of two or more magnetic layers separated by nonmagnetic spacers. A number of recent studies have focused on MR and related magnetoelectronic effects in thin-film bilayers comprising a ferromagnetic layer (FM) and a nonmagnetic layer with strong spin-orbit interaction (SOI), typically a heavy metal (HM) such as Pt, Ta, or W.[6-9] These studies are motivated by the rich phenomena that emerge from the interplay between SOI and magnetism, and their promising applications. In particular, it is now well established that electrical current in materials with strong SOI generates spin current due to the spin-Hall effect (SHE)[10-12] and/or the Rashba effect.[13,14] Injection of spin current into an adjacent FM layer can influence the static[6,7] and the dynamic[8,9] magnetization states of the latter. A particularly notable aspect of SOI-induced effects in FM/HM bilayers is their unidirectionality. For instance, for a given direction of the static magnetization, one polarity of the electrical current results in strong enhancement of magnetic fluctuations in the FM layer, while the opposite polarity results in their moderate suppression.[15] This dependence is reversed if the direction of the magnetization is reversed.

FM/HM bilayers also exhibit MR effects associated with SOI. The spin Hall magnetoresistance (SMR)[16,17] is believed to originate from the backflow of spin current from FM to HM. While the dependence of resistance on the magnetization direction for this effect differs from AMR, both are uniaxial – the resistance is symmetric with respect to the reversal of the magnetization or the current direction. In contrast, the recently discovered unidirectional spin Hall magnetoresistance (USMR)[18-23] is determined by the product $(\mathbf{j} \times \hat{\mathbf{z}})\mathbf{M}$, where $\mathbf{j}$ is the density of the electric current, $\mathbf{M}$ is the magnetization of the FM layer, and $\hat{\mathbf{z}}$ is the unit vector normal to the plane of the bilayer. Consequently, USMR changes sign when either the magnetization or the electric current is reversed.

The unidirectionality of magnetoresistance is particularly attractive for the magnetic memory technology, since it enables simple and robust detection of the direction of magnetization representing the information stored in the magnetic memory. In GMR- or TMR-based memory devices, this is achieved by utilizing a reference FM whose magnetization is pinned. The advantage of USMR is that the readout of information does not require additional pinned magnetic layers. Moreover, the magnetization direction of FM in FM/HM bilayers can be controllably reversed by the electric current[6,7]. Thus, such bilayers can be sufficient to implement a memory cell, in which spin current injection is utilized for writing the information, and USMR is utilized for reading. However, adequate understanding of USMR, necessary for the implementation of efficient memory devices utilizing this effect, has not yet been achieved. The proposed interpretations of USMR include the contribution to resistance of spin accumulation at the FM/NM interface[18] and spin-current driven excitation of magnons in the FM layer[21,23].

In this Letter, we report on electrical measurements of USMR, performed in conjunction with direct magneto-optical measurements of spin current-dependent magnon population, over a broad range of the driving currents and temperatures. Our magneto-optical Brillouin light scattering (BLS) measurements provide spectrally resolved information about magnon populations with an unprecedented sensitivity. We show that the current dependence of USMR closely follows the magnon density, and that both dependencies exhibit the same scaling over a large temperature range of 80-400 K. These findings demonstrate a close relationship between spin current-driven magnon generation and USMR. Furthermore, since BLS is sensitive predominantly to long-wavelength magnons, and the effects of spin current are magnon mode-dependent, our findings also indicate that USMR is dominated by the dipolar magnons, providing a possible route for the control and enhancement of this effect.

Our test devices are fabricated on the sapphire substrates, by using a combination of magnetron sputtering and electron-beam lithography. The devices consist of a 5 nm-thick and 2 μm in diameter $Ni_{80}Fe_{20}$ =Permalloy (Py) disk, fabricated on top of a 9 nm thick and 2.5 μm wide Pt microstrip (inset in Fig. 1(a)), which is electrically contacted by two 100 nm-thick Au leads. The two-probe resistance of the devices, including the leads, is about 50-80 Ω. Therefore, to achieve the accuracy of electronic resistance measurements of better than 1 mΩ, we perform averaging over $10^5$ measurements of every *R(I)* data point, with continuous zero check.

The magnon population in Py is detected using variable-temperature micro-focus BLS[24]. Single-frequency probing laser light with the wavelength of 532 nm is focused into a 450 nm spot at the center of the Py disk (inset in Fig. 1(b)), using a microscope objective lens with a large numeric aperture. The light scattered from magnons is collected by the same lens and analyzed, providing information about the spectral density of magnons, as described in more detail below. The sample is attached to the cold finger of a continuous-flow cryostat (Hires2, Oxford Instruments) equipped with an optical window, allowing control of the sample temperature in the range of 80-400 K with the accuracy of 1 K. The cryostat is placed in the gap of a permanent magnet producing a constant magnetic field *H*=690 Oe rotatable in the plane of the sample.

By rotating the static magnetic field *H* at a finite dc electric current *I*, we determine the quantity *ΔR(I)=R$_⊥$(I)-R$_{//}$(I)* characterizing the dependence of resistance on the magnetization orientation, as a function of current (symbols in Fig.1(a)). Here, *R$_{//}$(I)* and *R$_⊥$(I)* are the resistance values with the field directed parallel and perpendicular to the direction of current *I*, respectively, with positive direction of *I*, relative to *H*, defined as shown in the insets.

These data exhibit a clear asymmetry with respect to the current direction. At *I<0*, the value of *ΔR(I)* quadratically increases with increasing magnitude of current (dashed curve in Fig.1a). This dependence can be explained by a combination of SMR and AMR due to the rotation of magnetization by the Oersted field of the current. Additionally, since AMR decreases with increasing temperature, Joule heating can also contribute to the increase of *ΔR*. All these effects are symmetric with respect to the current reversal. Therefore, to the lowest order in current their contribution to *ΔR* is expected to be quadratic, in agreement with our measurements.

The increase of ΔR for *I>0* is considerably more significant than for *I<0*. To quantify this asymmetry, we define Δ$R_{USMR}(I)$=Δ*R(I)*-Δ*R(-I)*, as shown by the dimension line in Fig.1 (a). We note that the value of $R_{//}(I)$ is, within experimental error, independent of the current direction. Therefore, the observed asymmetry is associated entirely with the unidirectionality of $R_\perp$. Furthermore, the value of Δ$R_{USMR}$ is inverted when the direction of the transverse field is reversed, whereas reversal of the longitudinal field does not influence $R_{//}$. These symmetries are consistent with the prior studies of USMR,[18-23] and mirror those of SHE, indicating a close relationship between these two effects.

We now discuss the effects of spin current on the density of magnons characterized by BLS. The BLS intensity at the selected frequency *ω* is proportional to the spectral density of magnons *n(ω)*=$F_{BE}(ω)D´(ω)$, where $F_{BE}(ω)$ is the magnon occupation function described in equilibrium by the Bose-Einstein distribution, and *D´(ω)* is the density of magnon states weighted by the wavevector-dependent measurement sensitivity[25]. The latter determines the range *k*~0-$10^5$ cm$^{-1}$ of magnon wavevectors accessible to BLS. Accordingly, the spectral width of the BLS peaks is determined by the magnon dispersion within this range of wavevectors.

The BLS spectra, obtained with the transverse field, strongly depend on the driving current, as illustrated in Fig.1(b). This dependence is asymmetric with respect to the current direction, similarly to the USMR. For $I>0$, the intensity strongly increases with increasing current magnitude, but decreases for $I<0$. The behaviors are reversed when the applied field is reversed[15].

We characterize the asymmetry of the BLS spectra by using the quantity $\Delta G_{BLS}=G_{BLS}(I)-G_{BLS}(-I)$, where the current-dependent amplitude of the BLS peak $G_{BLS}(I)$ reflects the average spectral density of magnons in the range accessible to BLS. Figure 2 shows the current dependences of $\Delta R_{USMR}$ (up triangles) and $\Delta G_{BLS}$ (down triangles) determined at $T=$ 295 K and $T=$ 131 K, as indicated. The vertical scales were adjusted relative to one another, to provide the best matching between the two datasets. The results shown in Fig. 2 clearly demonstrate very similar current dependences of $\Delta R_{USMR}$ and $\Delta G_{BLS}$, as well as their similar scaling with temperature, suggesting a close relation between USMR and the current-dependent magnon density. We note that, in contrast to the result of Ref. 18, the observed dependences are nonlinear. Their extrapolation, as described below, diverges at a temperature-dependent current $I_C$ marked in Fig. 2 by the vertical dot lines. Such a divergence is expected for the effects of spin current on the magnon gas[15,25], due to the complete compensation of the natural magnetic damping.

We now analyze the effects of spin current on the magnon population. While the effects of spin current on magnetization dynamics are usually described using the deterministic Landau-Lifshitz-Gilbert-Slonczewski equation,[26] its effects on incoherent magnons are more naturally described by a statistical approach based on the Boltzmann equation.[15,27-30] Following this approach, we write the kinetic equation describing the evolution of dilute magnon gas:

$$\frac{\partial n}{\partial t} = \frac{\partial n}{\partial t}_{rel} + \frac{\partial n}{\partial t}_{sc}, \tag{1}$$

where $n=n(\omega,I)$ is the current-dependent spectral magnon density and $\frac{\partial n}{\partial t}_{rel} = -\frac{n-n_0}{\tau}$ is the relaxation term in relaxation-time approximation, whereas the term $\frac{\partial n}{\partial t}_{sc}$ describes the effect of spin current. Here, $n_0$ is the equilibrium spectral magnon density $n_0 = n(\omega, I = 0) = F_{BE}(\omega)D(\omega)$, where $F_{BE}(\omega)$ is the Bose-Einstein occupation function and $D(\omega)$ is the density of magnon states, and $\tau$ is the magnon relaxation time, which is related to the Gilbert relaxation parameter by $\tau=1/2\alpha\omega$. In the framework of the kinetic equation, the effect of spin current on the magnon gas can be described as stimulated emission of magnons by electron spin-flipping, at a rate proportional to the magnon population and to the difference $\Delta\mu_{\downarrow\uparrow}$ of the spin-dependent electrochemical potentials at the NM/FM interface, $\frac{\partial n}{\partial t}_{sc} = \varepsilon\Delta\mu_{\downarrow\uparrow}n$.[29] Neglecting the small effects of magnetization fluctuations on spin accumulation, $\Delta\mu_{\downarrow\uparrow}$ is proportional to the current, $\Delta\mu_{\downarrow\uparrow} = \beta I$, with the coefficient $\beta$ determined by the SHE efficiency and the geometry of the structure.

The stationary solution of Eq. (1) is

$$n(\omega, I) = \frac{n_0}{1-I/I_C}, \tag{2}$$

where $I_C=1/\varepsilon\beta\tau$ is the critical current, at which the magnon spectral density diverges. This result is consistent with the prior analyses of spin current effects, see e.g. Eq. (2) in Ref 15. The antisymmetric component of the dependence of magnon density on current is

$$n_A(\omega, I) = n(\omega, I) - n(\omega, -I) = n_0 \frac{2I/I_C}{1-(I/I_C)^2}. \tag{3}$$

The BLS intensity is proportional to the spectral magnon density,

$$\Delta G_{BLS}(I) = Bn_0 \frac{2I/I_C}{1-(I/I_C)^2}, \tag{4}$$

where the proportionality coefficient $B$ describes the sensitivity of the BLS apparatus.

The parameters B, $n_0$ and $I_C$ in Eq. (4) depend on the magnon frequency. In particular, $I_C \propto \omega$ since $\tau \propto 1/\omega$. However, since, on one hand, the effect of spin current on the frequency of the BLS peak is small and, on the other hand, the detected BLS peaks are rather narrow (Fig. 1(b)),[31] we neglect these dependencies in the analysis of the BLS peak amplitude. Solid curves in Fig. 2 show the results of the fitting of the experimental data for $\Delta G_{BLS}$ with Eq. (4), with $n_0B$ and $I_C$ used as the fitting parameters. A good agreement with the data confirms the validity of our model.

We further elucidate the relationship between USMR and the spin current-driven magnon population, by analyzing their temperature dependences. Since the measured $\Delta G_{BLS}$ and $\Delta R_{USMR}$ follow the same dependence on current, we approximate $\Delta R_{USMR}(I)$ by

$$\Delta R_{USMR}(I) = R_0 \frac{2I/I_C}{1-(I/I_C)^2}, \quad (5)$$

where $R_0$ is a scaling parameter given by the slope of the current dependence at small currents. If the USMR originates from the current-driven magnons, $R_0$ should be proportional to the equilibrium magnon density $n_0$, Eq.(3). The BLS intensity $G_{BLS}^0(T)$ at $I$=0 is also proportional to $n_0$. According to the Raleigh-Jeans law applicable to the degenerate low-frequency magnons accessible to BLS, $n_0$ is expected to depend linearly on temperature, and therefore so are $\Delta R_{USMR}$ and $G_{BLS}^0$.

Figure 3(a) shows the measured temperature dependences of $G_{BLS}^0$ (down triangles) and $R_0$ (up triangles), normalized by the corresponding values at room temperature. The BLS intensity varies linearly with temperature, in agreement with the Raleigh-Jeans law. At much lower temperatures, this dependence is expected to crossover to the $T^{3/2}$-law, consistent with the small positive intercept with the horizontal axis. The parameter $R_0(T)$ describing USMR follows precisely the same dependence, providing a strong experimental evidence for the magnon origin of USMR.

To further support this conclusion, we analyze the temperature dependences of the critical current $I_C(T)$ (Fig. 3b) obtained from the independent fitting of $\Delta R_{USMR}(I)$ and $\Delta G_{BLS}(I)$ with Eqs. (4) and (5), respectively. As seen from Fig. 3b, the two dependences coincide. The observed increase of the critical current with decreasing temperature is consistent with the reduction of spin Hall efficiency in Pt at low temperatures, as demonstrated in Ref. 32.

Although previous studies have already identified spin current-driven magnon generation as the possible origin of USMR,[21,23] a close relation between USMR and BLS intensity is surprising. Indeed, the BLS is sensitive to long wavelength magnons with wavevectors below $10^5$ cm$^{-1}$, and frequencies below 10-20 GHz, which occupy only a small part of the magnon Brillouin zone. In contrast, the dependence of resistance on the magnon population can be expected to be dominated by the large phase space of magnons with energies of the order of thermal energy, $f \approx \frac{k_B T}{h} \sim 2$ THz at $T$=100 K.[33] These much higher-frequency magnons are characterized by very different values of $n_0$ and $I_C$, and thus different dependences of populations on current and temperature, than the low-frequency dipolar magnons accessible to BLS.

The apparent inconsistency is resolved by considering the strongly non-equilibrium mechanism of spin current-driven magnon excitation. As discussed above, the efficiency of magnon excitation depends strongly on the magnon frequency, due to the frequency dependence of magnon relaxation time $\tau=1/2\alpha\omega$. In the framework of the Gilbert model, $\alpha$ is a constant, and $\tau \propto 1/\omega$, favoring the excitation of low-frequency magnons. Experimental studies show that at high magnon frequencies $\tau$ decreases with increasing frequency much faster than predicted by the Gilbert model,[34] leading to further suppression of the high-frequency magnon excitation. Additionally, it was recently shown[25] that the spin current generated by SHE drives the magnon gas into a quasi-equilibrium state described by the Bose-Einstein statistics with a positive chemical potential. Accordingly, spin current predominantly excites low-frequency magnons.

We experimentally test these arguments, by comparing the current dependences of the density of low-frequency magnons accessible to BLS with the total density of spin current-driven magnons. The former is characterized by the BLS peak amplitude $G_{BLS}(I)$, while the latter is obtained from the effective current-dependent magnetization $M_e(I)$. We determine the value of $M_e(I)$ from the current-dependent spectral position of the BLS peak, according to the Kittel formula for in-plane magnetized thin films[33]

$$\omega(I) = \gamma\sqrt{H(H + 4\pi M_e(I))}, \qquad (6)$$

where $\gamma$ is the gyromagnetic ratio, $H$ is the static magnetic field, and $M_e(I)=M_e(0) - 2\mu_B N(I)$ is the effective current-dependent static magnetization. Here, $N(I)$ is the total number of magnons excited by the spin current, and $\mu_B$ is the Bohr magneton. To characterize the asymmetry in the current dependence of the total magnon population, we define $\Delta M_e(I)=M_e(I)-M_e(-I)$. In Fig.4, we plot for two different temperatures $\Delta M_e(I)$ vs $\Delta G_{BLS}(I)$, modified by varying the driving current over the ranges shown in Fig.2. These data closely follow a linear dependence, indicating that the current-dependent variations of the total magnon population are likely dominated by the low-frequency magnons.

In summary, we utilized electronic and magneto-optical measurements of a Permalloy/Pt bilayer to analyze the current and the temperature dependences of the recently discovered unidirectional spin Hall magnetoresistance, as well as the spin current-induced variation of the low-frequency magnon density in the Py film. Our results demonstrate a close similarity in all the behaviors for these two phenomena, indicating their intimate connection. We believe that our findings will spur further progress in the understanding of electronic transport in spin-current driven magnetic systems, and contribute to the design of next-generation unidirectional spintronic and magnonic devices.

The authors acknowledge a fruitful discussion with Prof. P. Gambardella. This work was supported by the Deutsche Forschungsgemeinschaft, the NSF Grant Nos. ECCS-1509794, DMR-1504449, and ECCS-1804198, and the program Megagrant № 14.Z50.31.0025 of the Russian Ministry of Education and Science and by FASO of Russia (theme "Spin" No. AAAA-A18-118020290104-2).

**FIGURE CAPTIONS**

Fig. 1 (color online). (a) Current dependence of the $\Delta R=R_\perp-R_{//}$, as defined in the text, at $T=295$ K. Dashed line: quadratic fit of the $I<0$ data. Dotted and dimension lines illustrate the definition of unidirectional magnetoresistance $\Delta R_{USMR}$. (b) BLS spectra obtained at $T=295$ K, at the labeled values of current. Insets: schematics of the experiment and the definition of the positive current, for the shown direction of the transverse magnetic field.

Fig. 2 (color online) Current dependences of the unidirectional spin Hall magnetoresistance $\Delta R_{USMR}$ (up triangles), and of the unidirectional contribution to the BLS intensity $\Delta G_{BLS}$ (down triangles), obtained at $T=295$ K and $T=131$ K, as labeled. Solid lines show the results of the fitting of the experimental data for $\Delta G_{BLS}$ with Eq. (4). Vertical dashed lines mark the values of the critical current $I_C$ at which $\Delta G_{BLS}$ is extrapolated to diverge, as determined from the fitting.

Fig. 3 (color online) (a) Temperature dependencies of the USMR asymmetry parameter $R_0$ (up triangles), obtained by fitting the USMR data with Eq. (5), and of the BLS intensity $G^0_{BLS}$ obtained at $I=0$ (down triangles). Both data sets are normalized by their values at room temperature. The straight dashed line is obtained by a simultaneous linear fit of $R_0$ and $G^0_{BLS}$. (b) Temperature dependencies of the critical current $I_C$ obtained by fitting the USMR data with Eq. (5) (up triangles), and by fitting the data for $\Delta G_{BLS}$ with Eq. (4) (down triangles). The dashed line a guide for the eye.

Fig. 4 (color online) Unidirectional contribution $\Delta M_e(I)$ to the current-dependent reduction of the effective static magnetization vs the unidirectional contribution $\Delta G_{BLS}(I)$ to the BLS intensity, at $T$=295 K (diamonds) and $T$=131 K (squares).

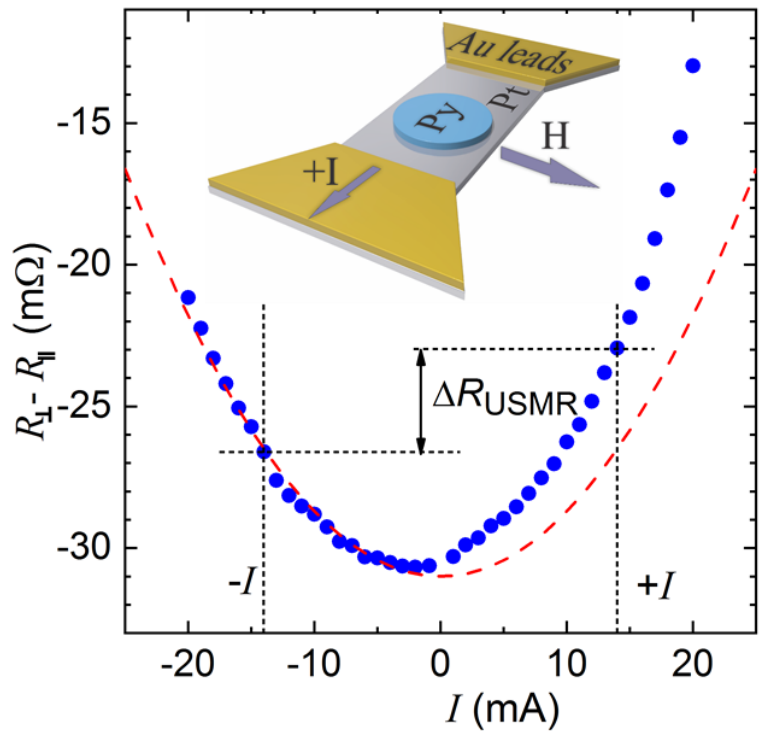

Fig.1a

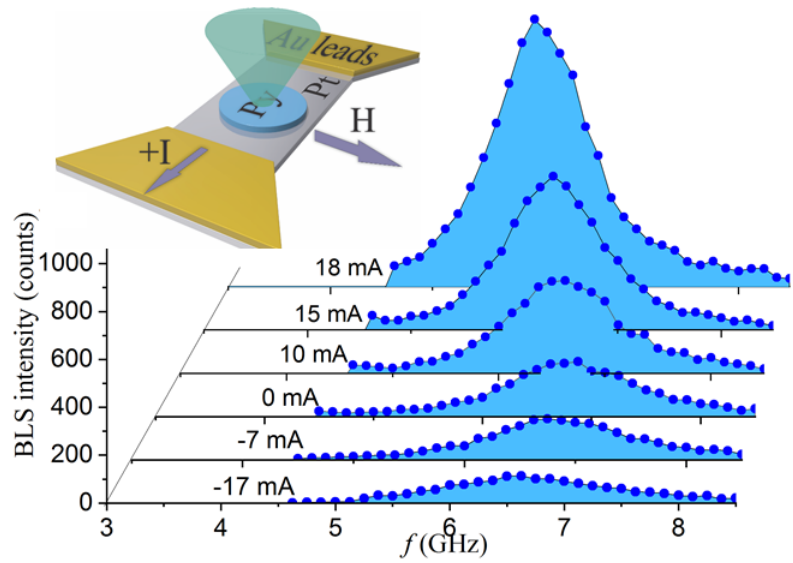

Fig. 1b

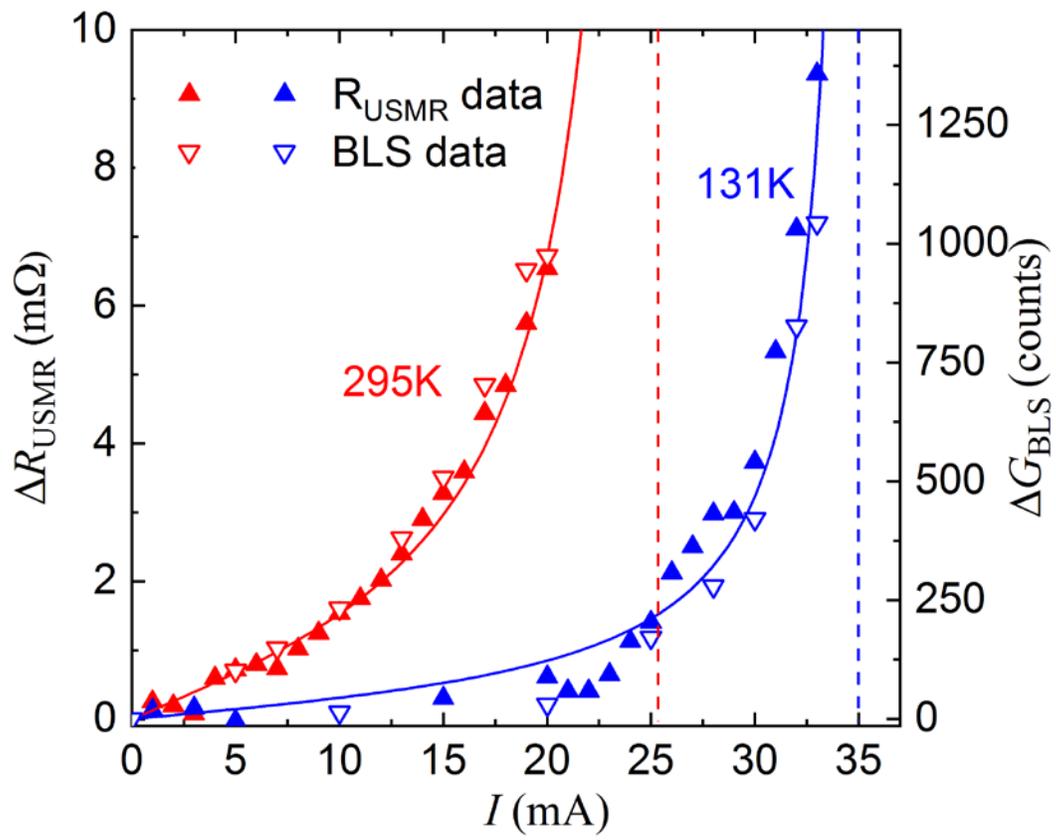

Fig. 2

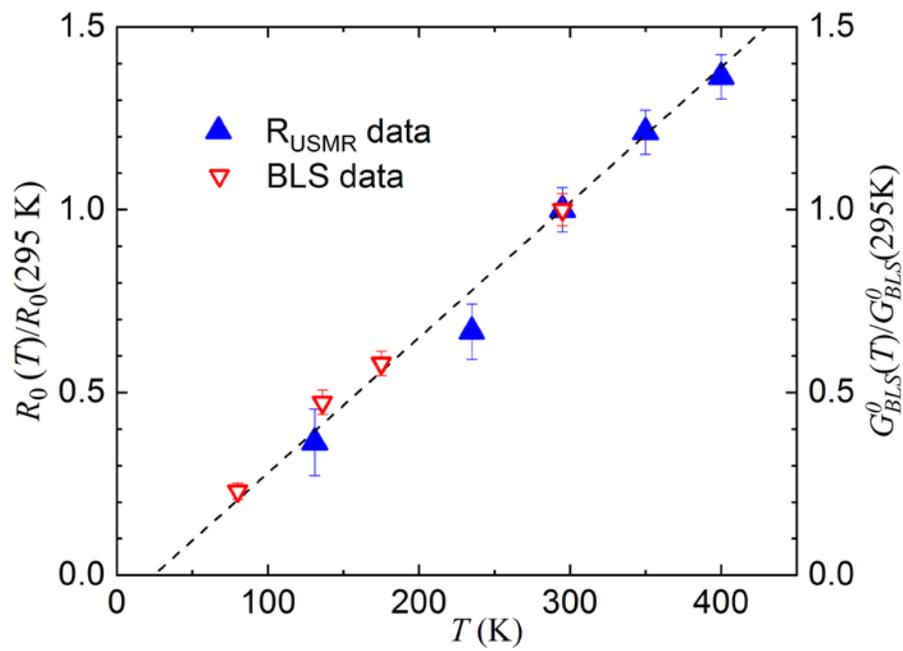

Fig. 3a

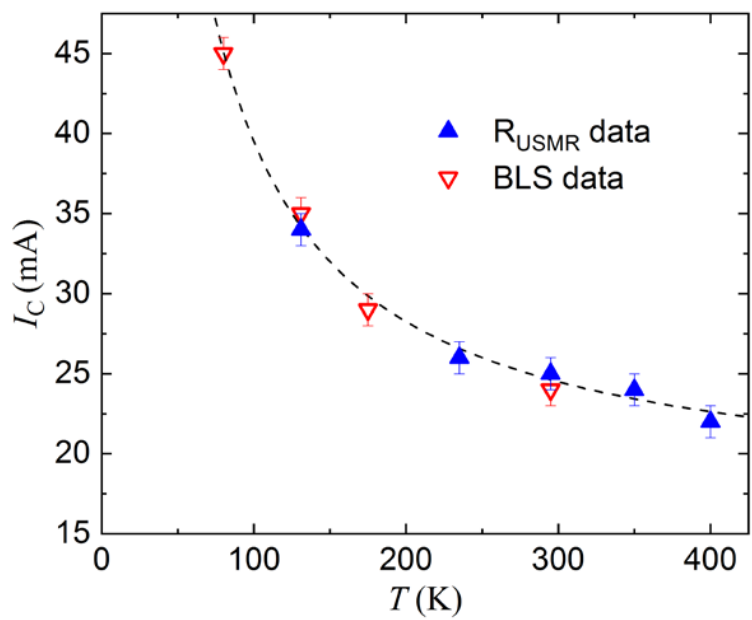

Fig. 3b

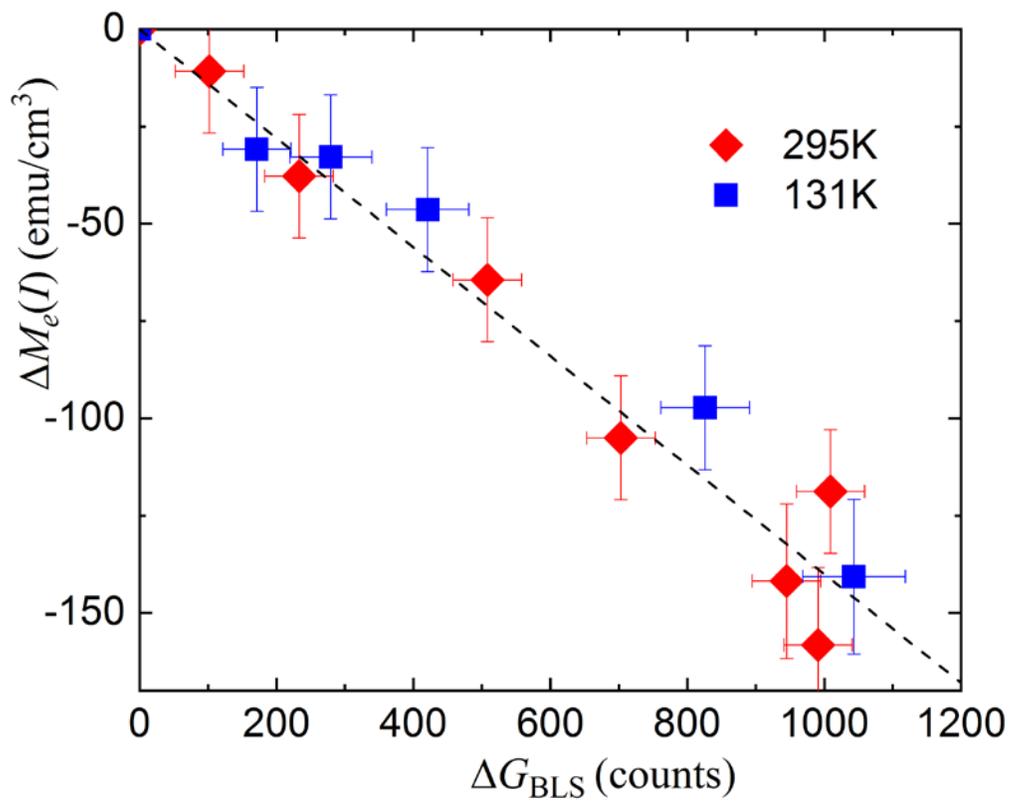

Fig. 4.